\documentclass[12pt]{article}

\usepackage{feynmp}
\usepackage{tikz-feynman}
\tikzfeynmanset{compat=1.1.0}
\usepackage{slashed}
\usepackage{subcaption}
\usepackage{amssymb,amsmath}

\DeclareMathOperator{\sgn}{sgn}

\begin{document}

\begin{titlepage}

\begin{flushright}
arXiv:2410.09273
\end{flushright}
\vskip 2.5cm

\begin{center}
{\Large \bf Bound States and Particle Production by\\
Breather-Type Background Field Configurations}
\end{center}

\vspace{1ex}

\begin{center}
{\large Abhishek Rout\footnote{{\tt arout@email.sc.edu}}
and Brett Altschul\footnote{{\tt altschul@mailbox.sc.edu}}}

\vspace{5mm}
{\sl Department of Physics and Astronomy} \\
{\sl University of South Carolina} \\
{\sl Columbia, SC 29208} \\
\end{center}

\vspace{2.5ex}

\medskip

\centerline {\bf Abstract}

\bigskip

We investigate the interaction of fermion fields with oscillating domain walls, inspired by breather-type
solutions of the sine-Gordon equation, a nonlinear system of fundamental importance.
Our study focuses on the fermionic bound states and particle
production induced by a time-dependent scalar background field. The fermions couple to two domain walls
undergoing harmonic motion, and we explore the resulting dynamics of the fermionic wave functions. We
demonstrate that while fermions initially form bound states around the domain walls, the
energy provided by the oscillatory
motion of the scalar field induces an outward flux of fermions and antifermions, leading to particle
production and eventual flux propagation toward spatial infinity. Through numerical simulations, we
observe that the fermion density exhibits quasiperiodic behavior, with partial recurrences of the bound state
configurations after each oscillation period. However, the fermion wave functions do not remain localized,
and over time, the density decreases as more particles escape the vicinity of the domain walls. Our results
highlight that the sine-Gordon-like breather background, when coupled non-supersymmetrically to
fermions, does not preserve
integrability or stability, with the oscillations driving a continuous energy transfer into the fermionic
modes. This study sheds light on the challenges of maintaining steady-state fermion solutions in
time-dependent topological backgrounds and offers insights into particle production mechanisms in nonlinear
dynamical systems with oscillating solitons.

\bigskip

\end{titlepage}

\newpage

\section{Introduction}

Particle production by extended classical structures is a topic that has been studied since the
earliest days of quantum mechanics. In nontrivial spacetime backgrounds, this has been an
incredibly fruitful line of inquiry, with the theoretical discoveries of the Unruh effect (for
an accelerating observer)~\cite{ref-unruh} and Hawking radiation (in a black hole spacetime)~\cite{ref-hawking}
having fundamentally changed our understanding of spacetime physics.
The creation of particles can be seen as a process of transitions between different Fock states of
a quantum theory embedded in an interesting classical background, and the relevant background can involve
fields other than the spacetime metric. In particular, the coupling of quantum-mechanical particles
particles to classical scalar and
vector fields can have many intriguing
consequences~\cite{ref-jackiw,ref-campbell,ref-riazi1,ref-riazi2,ref-bazeia1,ref-izquierdo1,ref-navarro-o}.
The nonlinear interactions responsible for both the shapes of solitary waves and the effective couplings
between multiple solitary
waves~\cite{ref-manton0,ref-gonzalez,ref-christov1,ref-christov2,ref-manton2,ref-manton1}
are significant topics of research by themselves. For instance, during a
sufficiently energetic collision of solitary wave pair in $1+1$ dimensions,
there may be a strongly inelastic bounce---in which
one or both of the solitary waves rebound back from the impact with internal modes or continuum
scalar field modes excited. Over long times, the behavior of repeatedly colliding and rebounding solitary
waves may display an intricate ``fractal'' dependence on the initial
conditions~\cite{ref-sugiyama,ref-anninos,ref-goodman1,ref-belendryasova1,ref-belendryasova2,ref-izquierdo2}.

However, when the background involves an extended structure built out of a bosonic field, the coupling of
one or more additional fermion fields can be
considered particularly interesting. Even without particle creation, the
coupling of fermions to a topologically nontrivial bosonic field configuration can lead to the
fascinating phenomenon of fermion fractionalization~\cite{ref-jackiw}.
The nontrivial topology of the background may produce
a potential for the fermion field whose spectrum is guaranteed to include a nondegenerate zero mode. When the
fermion field is quantized, with the Dirac sea of negative-energy states filled, the zero mode is assigned
partially to the Dirac sea and partially to the positive-energy part of the spectrum. The level in the
energy spectrum at which the fermion number $n_{F}$ vanishes
actually lies in the middle of the zero mode~\cite{ref-su}.
Since the zero mode cannot ever actually be only partially filled, the real Fock states
(with the mode filled or unfilled) consequently have non-integer values of
$n_{F}$~\cite{ref-kivelson,ref-frishman}.

The situation may be even more complicated when there are multiple solitary wave structures present in
the background. In the presence of multiple stationary solitons, there are typically multiple zero-energy
(or almost-zero-energy) modes as part of the fermionic spectrum. When the solitary wave features are in
motion relative to one-another, this is further augmented by the possibility of fermion or antifermion
particle creation~\cite{ref-hirata}, and it is a type of scenario like this that will be the focus of this paper.
This is part of an ongoing research program~\cite{ref-karki4}, which has so far examined the fermion spectrum
in backgrounds with two or three solitary waves. However, one of the challenges to doing calculations
in a theory like this---in which the energy differences between fermion states bound to bosonic
solitary wave may be exponentially small---is that the background may not be known precisely. Solitary waves
are a feature of nonlinearly coupled classical fields, and in most cases analytic solutions involving multiple
topological excitations are either unknown or extraordinarily complicated and impossible to express concisely.
Small discrepancies between the precise energies of the background field configurations and the approximate
forms used for calculational expediency may become a problem when trying to study the very small energy
differences between neighboring fermion states.

In this paper, we shall look at fermion solutions in a particular type of background, inspired by the
sine-Gordon equation in $1+1$ dimensions, since the sine-Gordon equation has time-dependent,
multiple-solitons solutions
that are known analytically. The focus will continue to be on the manifold of almost-zero-energy states that
generalize the Jackiw-Rebbi zero modes; our work represents new fundamental advances in this area of
investigation. In section~\ref{sec-breathers}, we shall review the
sine-Gordon equation and its breather-type soliton solutions. The
consideration of a breather background was motivated by the hope that the
breathers, which are bound soliton-antisoliton states with periodic time dependence, might support fermion
bound states that were similarly periodic and for which analytic solutions might be even be found. However,
solving the Dirac equation in this kind of time-dependent background still turns out to be too difficult to
do analytically, so in section~\ref{sec-Dirac-eq}, we introduce another kind of bosonic background,
inspired by the sine-Gordon breather but involving step functions that make the fermionic theory amenable to
additional analytic solutions methods. In section~\ref{sec-fourier} we show, using a Fourier decomposition
of the fermionic modes, that for neither the step function background, nor for the true sine-Gordon breather
background, can there actually be steady-state fermion solutions. There is an unavoidable outward flux in
the fermion and antifermion nodes, driven energetically by the oscillations of the background scalar field.
This is an entirely new result.
In section~\ref{sec-computations}, we discuss the computational challenges this poses and present numerical
results for how fermionic states that start off closely bound to the solitary waves in the background produce
wave of flux escaping to spatial infinity. Section~\ref{sec-concl} gives our final conclusions.

\section{Breather Solitons}

\label{sec-breathers}

The sine-Gordon theory for a scalar field in $1+1$ dimensions has the Lagrange density
\begin{equation}
\mathcal{L}_{\phi}=\frac{1}{2}(\partial^{\mu}\phi)(\partial_{\mu}\phi)-\frac{1}{\pi}[1-\cos(\pi\phi)].
\end{equation}
The theory this describes is extremely special, having degenerate vacua at all $\phi=n$ and solitary wave
solutions that interpolate between them.  These solitary waves are quite unusual, since they
can pass through one-another with no net deformations or radiation once they are well separated again.
The solitary waves are thus actually ``solitons'' in the oldest and strongest sense of that term.
The theory is integrable, a particularly unusual property in a relativistic theory and one that is
closely connected to the solitary wave structure. The schematic (but
nonrigorous) argument that the presence of true
solitons means that there must be an infinite number of conservation laws is fairly straightforward
Suppose that $N$ solitons are moving along $x$-axis; any $N\geq0$ is possible. They can all have different
speeds, so that as $t\rightarrow\infty$ the separations between the solitary waves
will become arbitrarily large, and thus the fraction of the total energy that remains tied up in the
interactions between the solitons shrinks arbitrarily small. Each soliton is essentially an isolated system
with conserved internal and kinetic energies of its own; and 
since $N$ may be arbitrarily large, there must likewise be an unlimited number of these conservation laws.
Such localized single-soliton energies are quite different from the usual 
way in which the infinite family of sine-Gordon conservation laws are normally expressed.
However, the motivation for this argument was simply to show the deep connection between the
existence of many-soliton solutions for $\phi$ and existence of an infinite number of conservation laws.
(Related analyses of integrable equations in $2+1$ dimensions are also currently an active area of
study~\cite{ref-wazwaz,ref-ma,ref-gao1,ref-gao2}.)

A very natural question to ask is how many of the special properties of the sine-Gordon theory persist
when the bosonic theory is coupled to a Dirac fermion field. Solutions with single solitary waves coupled
to the fermions are well understood, with the unusual observation being the presence of zero-energy
bound state modes and the surprising consequence that the fermion number actually has fractional
eigenvalues~\cite{ref-jackiw}---among other anomalous
features~\cite{ref-rajaraman,ref-shifman1,ref-vachaspati,ref-amado}.
Studying fermions in scalar backgrounds composed of multiple solitary wave
is a complicated subject, however. Normally, it is only amenable to approximate analyses---whether analytical
or numerical~\cite{ref-karki4,ref-graham2,ref-goldhaber,ref-brihaye,ref-perapechka}.
However, one might hope that the simplicity of the sine-Gordon system would make it more
straightforward to locate complete analytic solutions of the coupled boson-fermion system. Unfortunately,
this does not
seem to be the case, and, in fact, the first step in our analysis will be to replace the sine-Gordon breather
background with a seemingly artificial approximation. Nonetheless, we shall find that our results can reveal
quite a bit about the general structure of the Dirac theory in a time-dependent oscillatory topological
background.

Beyond scattering states, in which two sine-Gordon solitons collide---either passing through each other or
rebounding---there are also bound states of the solitons. In particular,
the sine-Gordon equation has breather solutions (normalized so that they interpolate between
vacuum states at $\phi=-1$, $0$, and $1$ and oscillate with period $2\pi\sqrt{1+b^{2}}/b$),
\begin{equation}
\label{eq-phib}
\phi_{b}(x,t)=\frac{4}{\pi}\tan^{-1}\left[\frac{\sin\left(bt/\sqrt{1+b^{2}}\right)}
{b\cosh\left(x/\sqrt{1+b^{2}}\right)}\right];
\end{equation}
and the fermion Lagrange density in the presence of the sine-Gordon background field $\phi$ is
\begin{equation}
\label{eq-Lfermion}
\mathcal{L}_{\psi}=\bar{\psi}\left[i\slashed{\partial}+g\cos\left(\frac{\pi}{2}\phi\right)\right]\psi.
\end{equation}
This is the ``natural'' form for the coupling of the Dirac field to the scalar, because with a
properly chosen value of $g$, the action becomes supersymmetric. In that case, the
interactions in the boson and fermion Lagrange densities
each descend from a ``superpotential'' $W(\phi)=\sin(\pi\phi/2)$, since $(1-\cos\pi\phi)\propto W^{2}$
and $\cos(\pi\phi/2)\propto dW/d\phi$. (However, Ref.~\cite{ref-brihaye} used
an ordinary Yukawa coupling even when the bosons obeyed a sine-Gordon equation.)
Perhaps ironically, however, we do not actually expect our general conclusions to apply in the precisely
supersymmetric sine-Gordon model, which remains integrable---like the purely bosonic
model~\cite{ref-ferrara,ref-bajnok}. The
existence of an infinite number of degrees of freedom can qualitatively change the behavior of the system.

\section{Dirac Equation in the Kink-Antikink Background}

\label{sec-Dirac-eq}

Unfortunately, no analytic solutions to the Dirac equation in the breather background have been forthcoming.
So instead of the anharmonically-oscillating
sine-Gordon breather solution, we shall take two
step function domain walls undergoing simple harmonic motion back and forth through one-another,
with amplitude $v\geq0$ and frequency $\omega$,
\begin{equation}
\label{eq-kinks}
\phi(x,t)=\sgn\left(x-v\cos\omega t\right)-\sgn\left(x+v\cos\omega t\right).
\end{equation}
[We shall refer to the first term on the right-hand side of \eqref{eq-kinks} as the ``kink,'' which
is located at positive $x$ at the initial time $t=0$, and the second term as the ``antikink.'']
In this case, the scalar potential appearing in the Dirac equation
$\left[i\slashed{\partial}+V(x,t)\right]\psi=0$ is
\begin{equation}
\label{eq-V-sgn}
V(x,t)=g\sgn\left(x^{2}-v^{2}\cos^{2}\omega t\right),
\end{equation}
so with the Dirac matrix representation
$\gamma_{0}=\sigma_{1}$ and $\gamma_{1}=i\sigma_{3}$, the $2\times 2$ matrix form of the Dirac
equation becomes
\begin{equation}
\label{eq-matrixDirac}
\left[
\begin{array}{cc}
\partial_{x}+g\sgn\left(x^{2}-v^{2}\cos^{2}\omega t\right) & i\partial_{t} \\
i\partial_{t} & -\partial_{x}+g\sgn\left(x^{2}-v^{2}\cos^{2}\omega t\right)
\end{array}
\right]
\left[
\begin{array}{c}
\psi_{1} \\
\psi_{2}
\end{array}
\right]=\left[
\begin{array}{c}
0 \\
0
\end{array}
\right].
\end{equation}

With a Yukawa coupling $-g\phi\bar{\psi}\psi$ (as would be natural in a $\phi^{4}$ theory)
instead of the cosine coupling term in \eqref{eq-Lfermion}, a kink would produce an attractive potential
for the upper component $\psi_{1}$ of the Dirac wave function, and repulsive potential for the
lower component $\psi_{2}$. The zero-mode wave function for a bound fermion attached to a single
stationary signum function located at $x_{K}$ would be (assuming a coupling strength $g>0$)
\begin{equation}
\left[
\begin{array}{c}
\psi_{1} \\
\psi_{2}
\end{array}
\right]=
\left[\frac{g4^{g}\Gamma(g+\frac{1}{2})}{\sqrt{\pi}\,\Gamma(g+1)}\right]^{1/2}\left[
\begin{array}{c}
e^{-g|x-x_{K}|} \\ 0
\end{array}
\right].
\end{equation}
Similarly, the zero mode attached to an isolated antikink has only its lower component nonzero.
The specific exponential wave functions are particular to the infinitely narrow domain walls, but the
general property that one component of the zero-mode wave function vanishes is generic, applying in
any stationary background for which $V(-\infty)$ and $V(+\infty)$ have opposite signs.

In contrast, with the interaction \eqref{eq-Lfermion}, what determines which component of a zero-mode
wave function attached to a solitary wave is nonzero is not whether $\phi$ is increasing (kink) or
decreasing (antikink). Instead, it depends on whether the cosine function
in $\mathcal{L}_{\psi}$ is increasing or decreasing across
the solitary wave. That is determined by whether a particular solitary wave interpolates from an even to an odd
vacuum value---meaning from $V(-\infty)=2n$ to $V(+\infty)=2n\pm1$---or from an odd vacuum to an even one.
A kink interpolating between the degenerate vacua at $\phi=0$ and $\phi=1$ acts the same way on the
Dirac wave function as an antikink interpolating between $\phi=0$ and $-1$.

In the presence of a well-separated kink-antikink pair, there are two almost-zero-energy fermion modes,
with even and odd behaviors under parity. When the domain walls are far apart, the corresponding fermion wave
function are well approximated by symmetric and antisymmetric linear combinations of a purely-$\psi_{1}$
bound state wave function tightly localized around one domain wall's location and a purely-$\psi_{2}$ function
localized around the other solitary wave. In this regime, the two fermion states are approximately equally
displaced above and below zero energy. In contrast, when the kink and antikink draw close together,
so that $v|\cos\omega t|\lesssim g^{-1}$, the wave functions become much more complicated. Both $\psi_{1}$
and $\psi_{2}$ are appreciable in the neighborhood of $x=0$, and their energy shifts are no longer
equal and opposite.
Moreover, even though the kink and antikink pass through one-another, the almost-zero-energy
modes do not follow along. After their episodes of complicated mixing during each interval in which
$v|\cos\omega t|\lesssim g^{-1}$, the $\psi_{1}$-dominated and $\psi_{2}$-dominated portions of the
full wave function do not pass through each other but instead rebound, switching their
locations of attachments from kink to antikink and vice versa. The wave function $\psi$ is periodic
with period $\pi/\omega$, like $V$, rather than the $2\pi/\omega$ periodicity of $\phi$ itself.
Moreover, the evolution of the system during the collision period cannot be adiabatic; although the
adiabatic approximation may be excellent when the kink and antikink are widely separated, the field
backgrounds \eqref{eq-phib} or \eqref{eq-kinks} both pass through the state $\phi=0$---which
is a background in which the discrete modes of the
instantaneous Hamiltonian merge into a massless continuum that extends through $E=0$ with no gap.

Using a prime to denote a derivative with respect to $x$ and a dot one with respect to $t$,
the coupled partial differential equations \eqref{eq-matrixDirac} for the upper and lower components are
\begin{eqnarray}
\label{eq-psi1'}
\psi_{1}'+V\psi_{1}+i\dot{\psi}_{2} & = & 0 \\
\label{eq-psi2'}
-\psi_{2}'+V\psi_{2}+i\dot{\psi}_{1} & = & 0.
\end{eqnarray}
Unlike the system of time-independent energy eigenvalues, these cannot be fully separated, because of
the time dependence of the potential $V$. From the general forms (\ref{eq-psi1'}--\ref{eq-psi2'}), we may
we may differentiate \eqref{eq-psi1'} spatially and use the time derivative
of \eqref{eq-psi2'} to eliminate $\dot{\psi}_{2}'$. However,
since \eqref{eq-psi2'} involves both $\psi_{2}'$ and $\psi_{2}$,
this substitution introduces terms with a $V\dot{\psi}_{2}$ and $\dot{V}\psi_{2}$;
the first of these may be eliminated in turn by one
more substitution of the (undifferentiated) \eqref{eq-psi1'}, but the second one, which appears because
of the time dependence of $V$, cannot. This leaves the two equations still unavoidably coupled.
The final resulting equations for both components of $\psi$ are
\begin{eqnarray}
\label{eq-psi1''}
\ddot{\psi_{1}}-\psi_{1}''+(V^{2}-V')\psi_{1}-i\dot{V}\psi_{2} & = & 0 \\
\label{eq-psi2''}
\ddot{\psi_{2}}-\psi_{2}''+(V^{2}+V')\psi_{2}-i\dot{V}\psi_{1} & = & 0.
\end{eqnarray}
For the particular $V(x,t)=g\sgn\left(x^{2}-v^{2}\cos^{2}\omega t\right)$, this simplifies even further,
since $V^{2}=g^{2}$, so that except at the locations of the kink and antikink, each equation simply has
the form of a free massive Klein-Gordon equation, $(\Box+g^{2})\psi_{j}=0$. Moreover, since
$\dot{V}=-v\omega\sin(\omega t)V'$, we have
\begin{equation}
(\Box+g^{2})\psi_{j}+(-1)^{j}2g\left[\delta(x-v|\cos\omega t|)-\delta(x+v|\cos\omega t|)\right]
\left[\psi_{j}+(-1)^{j}iv\omega\sin(\omega t)\psi_{3-j}\right]=0.
\end{equation}
The fact that the solitary wave located at positive $x=v|\cos\omega t|$ is always the one
that is attractive to the
$\psi_{1}$ component and repulsive to $\psi_{2}$, regardless of whether the solitary wave is a kink or
antikink is again evident in this equation.
Using a Klein-Gordon Green's function, these two coupled partial differential equations could be
converted into
two coupled double-integral equations; then the $\delta$-functions would eliminate one of the two
integrations, although it is not clear whether this could be useful in practice.

\section{Fourier Modes of the Dirac Equation}

\label{sec-fourier}

Since the potential $V$ is periodic with period $\pi/\omega$,
any steady-state solution ought to be as well, so we can try
representing $\psi$ as a Fourier series,
\begin{equation}
\psi(x,t)=\sum_{n=-\infty}^{\infty}
\left[
\begin{array}{c}
\psi_{1n}(x)e^{-2in\omega t} \\
\psi_{2n}(x)e^{-2in\omega t}
\end{array}
\right].
\end{equation}
However, this does not, on its own, split up the Dirac equation into a set of linear ordinary
differential equations for the $\psi_{jn}$ Fourier components, because the
left-hand side of \eqref{eq-matrixDirac} is still time dependent. However, we may also Fourier
transform the step function $V$. For each $x$,
\begin{equation}
V(x,t)=g\left[\frac{1}{2}c_{0}(x)+\sum_{m=1}^{\infty}c_{m}(x)\cos(2m\omega t)\right],
\end{equation}
so that, for $m\neq0$,
\begin{equation}
\int_{0}^{\pi/\omega}dt\, g\sgn\left(x^{2}-v^{2}\cos^{2}\omega t\right)\cos(2m\omega t)=
g\sum_{m'=1}^{\infty}c_{m'}(x)\int_{0}^{\pi/\omega}dt\, \cos(2m'\omega t)\cos(2m\omega t),
\end{equation}
in which the right-hand side is simply $(\pi g/2\omega)c_{m}(x)$. The signum function splits the
left-hand side into two integrals,
\begin{equation}
\label{eq-split-t0}
g\int_{0}^{t_{0}}dt\,[-\cos(2m\omega t)]+g\int_{t_{0}}^{\pi/\omega}dt\,\cos(2m\omega t)=-\frac{2g}{2m\omega}
\sin(2m\omega t_{0}),
\end{equation}
where $t_{0}$ is the time at which the kink is located at position $x$,
\begin{equation}
t_{0}=\left\{
\begin{array}{ll}
0, & |x|>v \\
\frac{1}{\omega}\cos^{-1}\frac{|x|}{v}, & |x|<v
\end{array}
\right..
\end{equation}
This makes the Fourier coefficients
\begin{equation}
c_{m>0}(x)=\left\{
\begin{array}{ll}
0, & |x|>v \\
-\frac{2}{m\pi}\sin\left(2m\cos^{-1}\frac{|x|}{v}\right), & |x|<v
\end{array}
\right..
\end{equation}
For the $m=0$ case, the left-hand side of \eqref{eq-split-t0} is simply $g[(\pi/\omega)-2t_{0}]$,
and so $c_{0}(x)$, which is just twice the time average of $V$ at the location $x$, is
\begin{equation}
c_{0}(x)=\left\{
\begin{array}{ll}
2, & |x|>v \\
2\left(1-\frac{4}{\pi}\cos^{-1}\frac{|x|}{v}\right), & |x|<v
\end{array}
\right..
\end{equation}

%making the ordinary differential equations for each Fourier mode
%\begin{eqnarray}
%\label{eq-psi1n'}
%\psi_{1n}'+g\sgn\left(x^{2}-v^{2}\cos^{2}\omega t\right)\psi_{1n}+n\omega\psi_{2n} & = & 0 \\
%\label{eq-psi2n'}
%-\psi_{2n}'+g\sgn\left(x^{2}-v^{2}\cos^{2}\omega t\right)\psi_{2n}+n\omega\psi_{1n} & = & 0.
%\end{eqnarray}

The fact that $V$ is time dependent means that all the spatial differential equations for the
$\psi_{jn}$ are coupled together, so solving the system is still a challenge. In fact, inserting
the Fourier expansion of $V$ into (\ref{eq-psi1'}--\ref{eq-psi2'}) produces an infinite number
of coupled first-order linear differential equations. However,
when $|x|$ is larger than the domain walls' oscillation amplitude $v$, then the signum function is 1
at all times (all $c_{m>0}=0$), and the equations for the $\psi_{jn}$ decouple,
\begin{eqnarray}
\psi_{1n}'+g\psi_{1n}+2n\omega\psi_{2n} & = & 0 \\
-\psi_{2n}'+g\psi_{2n}+2n\omega\psi_{1n} & = & 0,
\end{eqnarray}\
and have straightforward solutions,
\begin{eqnarray}
\psi_{1n} & = & A_{n}e^{i\sqrt{4n^{2}\omega^{2}-g^{2}}\,x}+B_{n}e^{-i\sqrt{4n^{2}\omega^{2}-g^{2}}\,x} \\
\psi_{2n} & = & -A_{n}\frac{g+i\sqrt{4n^{2}\omega^{2}-g^{2}}}{n\omega}e^{i\sqrt{4n^{2}\omega^{2}-g^{2}}\,x}
-B_{n}\frac{g-i\sqrt{4n^{2}\omega^{2}-g^{2}}}{n\omega}e^{-i\sqrt{4n^{2}\omega^{2}-g^{2}}\,x}\,\,\,\,\,\,\,
\end{eqnarray}
as complex exponentials.
For large $|g|>2|n|\omega$, these solutions look quite reasonable as Fourier components of a wave
function; either $A_{n}$ or $B_{n}$ will vanish so that there is no exponential growth at infinity,
and the asymptotic spatial behavior is $\psi_{jn}\sim e^{-\sqrt{g^{2}-4n^{2}\omega^{2}}\,|x|}$.

However, it may initially appear that things go awry for large $|n|$. It seems extremely unlikely that
all the Fourier components will be identically zero for $|x|>v$ beyond some maximum $|n|$. Yet if
arbitrarily-high-frequency terms are present in the Fourier expansion, then it is clear that beyond
$|n|>|g|/2\omega$, the $\psi_{jn}$ terms will not be exponentially decaying as $|x|\rightarrow\infty$ but
will instead exhibit spatial oscillations.

The resolution to this puzzle comes from the fact that the Dirac wave function does not have a true
single-particle
probabilistic interpretation and need not be normalizable. There is the well-known Klein
paradox~\cite{ref-klein1,ref-sauter1},
in which a repulsive external potential of height greater than $2m$ can have oscillatory
solutions where the potential is highest. In that situation, the potential can supply enough energy to
create real fermion-antifermion pairs, and the presence of oscillatory solutions in the classical
forbidden region is a manifestation of this effect. In the situation we are examining here, there is
likewise an external source of energy that may be transferred to the fermion field via particle creation.
Regardless of how small $g$ and $v$ are, the back-and-forth oscillations of the domain wall pair can
eventually contribute enough energy to create real outgoing particles, because the oscillations
have been taken to continue eternally. Over time, it is possible for the Dirac field to absorb $|n|$ quanta
from the coherent domain wall oscillations, regardless of how large $|n|$ is.

So the solutions of the Dirac equation in the time-dependent background cannot be expected to show only
exponential damping at large $|x|$; there is also the possibility for an outflow of particles. For a mode to
have enough energy to represent escaping continuum particles, it needs to represent transfer of enough
quanta of energy $\omega$ to create the massive particles that escape to infinity; but this is
precisely those modes for which $\sqrt{4n^{2}\omega^{2}-g^{2}}$ is real! Setting boundary conditions
so that there are only outgoing quanta (no incoming modes excited) and including the time dependences of
the components, the long-range behavior is
\begin{equation}
\label{eq-large-x}
\psi_{jn}e^{-2in\omega t}\sim e^{2in\omega\left(\sqrt{1-g^{2}/4n^{2}\omega^{2}}\,|x|-t\right)}.
\end{equation}

The observation that a steady-state, $(\pi/\omega)$-periodic solution of the Dirac-equation
necessarily involves a outgoing flux of quanta from the kink-antikink region suggests that
the one of the original reasons for focusing especially on the sine-Gordon breather system was
misaimed.  The purely bosonic sine-Gordon theory is integrable, with an infinite number of conserved
quantities. Recall, the solitary waves are true ``solitons'' in the strong sense, in that
while they exert net forces on one-another, when the collide they will pass through one-another and
as they separate to great distances, return to their original shapes. In particular, while there are
the breather solutions, in which a kink and an antikink form a bound state, never will a kink and
an antikink annihilate into non-topological low-amplitude
radiation excitations. It was hoped that including
appropriately coupled fermions in the theory (perhaps in the version of the boson-fermion theory
which differs from the supersymmetric theory only in the magnitude of the coupling constant)
might preserve the integrability of the theory and the absolute stability of individual solitary
waves. However, the
argument for the structure \eqref{eq-large-x} for the large-$n$ Fourier modes' behavior in the far field
does not depend in any fundamental way on the specific step function approximation for the $\phi$ background;
the reasoning would apply equally well with \eqref{eq-phib} as with \eqref{eq-kinks}.
Any one of the large-$n$ modes should generally be excited, unless there is a conservation law that prohibits
such an excitation. This is why the choice of coupling that makes the
fermion-coupled sine-Gordon model truly supersymmetric
differs from all other values. The supersymmetric version has an infinite number of conservation laws and
is thus capable of preventing the excitation of the infinite number of outward-propagating modes and
preserving perpetual breathers.
However, the fact that the solutions of the Dirac equation in a breather-like background
otherwise involve outward
net flux suggests that any hope that the most general sine-Gordon theory coupled to fermions
might remain fully integrable appears to be untenable.

\section{Computational Complications and Results}

\label{sec-computations}

The fact that fermionic modes can propagate outside the kink-antikink region (completely freely
in the square wave model, nearly freely in the more realistic sine-Gordon theory) poses numerical
as well as analytical challenges. Because of the greater complexity of the sine-Gordon breather system,
we shall continue to focus, as we move to numerical calculations, on the model with step function
solitons. We shall also neglect the back-reaction of the fermions on the boson field, treating
$\phi$ as a unmodified oscillating background. This is a natural approximation when the coupling $g$ is
small, and is in any case necessary when working with the signum function domain walls, whose dynamical
equations would involve singular quantities.

An algorithm for integrating the Dirac equation is going to
encounter difficulties at the boundaries of the integration domain, because the solutions of the
unmodified Dirac equation may be dominated by convective behavior---which will typically involve
energies modes propagating inward from infinity as well as outward toward infinity. This is obviously
contrary to the radiation boundary conditions we would like to impose.

If the initial Dirac wave function is well localized around the soliton-containing region, then at small
times the numerical solution in the radiation zone will be dominated, as desired, by the
outgoing fermion and antifermion modes. However, if the Dirac field is not subject to unphysical
conditions (such as the stong spatial damping described below)
at or near the ends of the integration domain, then when the outgoing excitations start to
impinge on the exterior boundaries, the algorithm will suffer from an unstable explosion of unphysical
incoming mode excitations. A small numerical error will introduce an admixture of incoming waves into the
solution, with the error due to these spurious modes increasing with time.
The field itself will begin to blow up at the boundary, and the inward
propagation will rapidly wreck the calculated solution.

\begin{figure}
\centering
\includegraphics[angle=0,width=14cm]{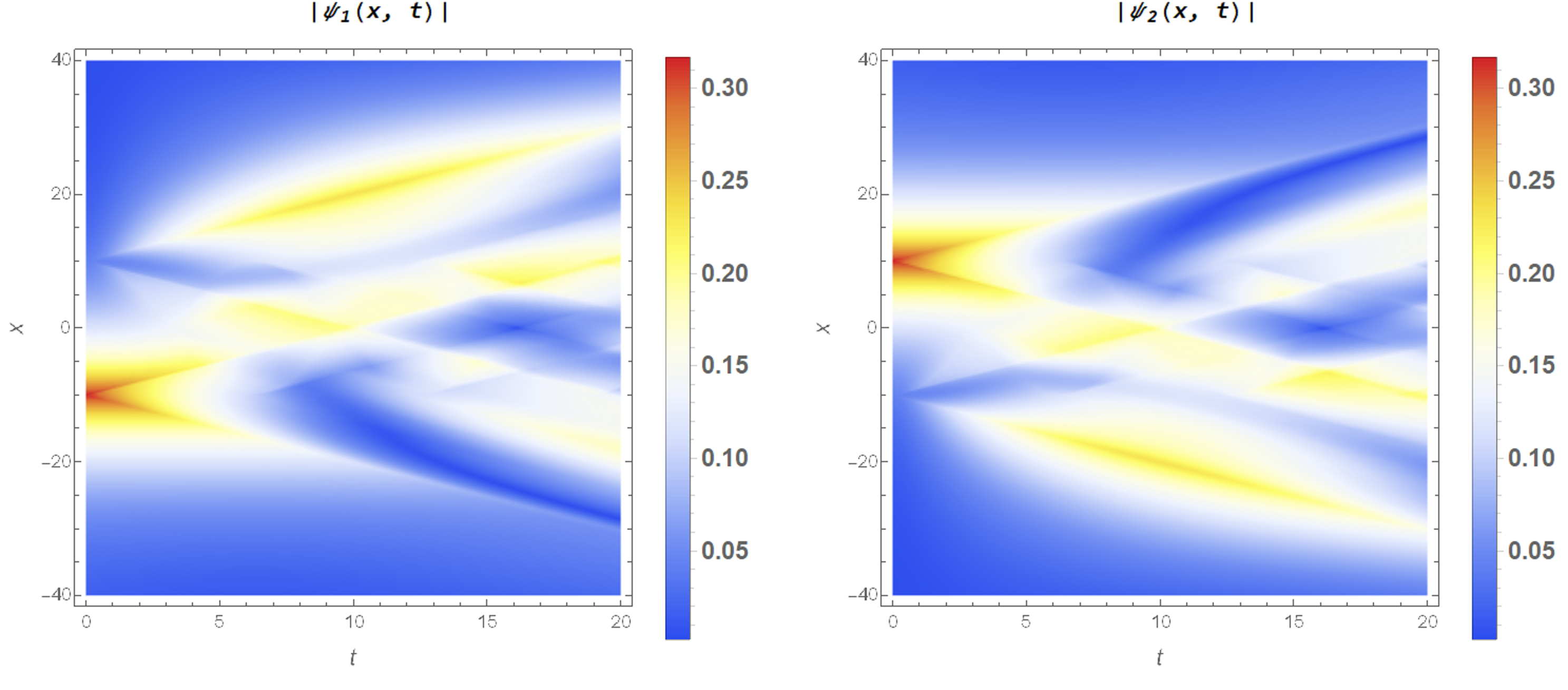}
\caption{The upper (left) and lower (right) components of the time-evolving Dirac wave function,
plotted in terms of position $x$ and time $t$ for the coupling value $g=0.10$, domain wall oscillation
amplitude $v=10$, and frequency $\omega=0.50$.
The starting $\psi$ is a symmetric bound state of the instantaneous
$V(x,t=0)$.}
\label{fig-abs0.10}
\end{figure}

There might seem to be an obvious solution to the instabilities at the left and right boundaries---applying
strong damping to the equation in the vicinities of the boundaries. Indeed, some of the built-in integration
algorithms in Mathematica software are ``clever'' enough to make this suggestion. Applying such an
unphysical modification to the standard Dirac equation does indeed improve the reliability of
integration algorithms for a certain amount of time. However, over longer periods, this method is not
satisfactory either. The problem arises from the fact that the method is similar to pegging the value
of the Dirac field to zero at the ends of the integration region. As a result, over longer times,
the solution eventually degrades into what is essentially a driven (by the oscillating domain walls)
standing wave, rebounding back and forth between the two ends of the integration interval.

\begin{figure}
\centering
\includegraphics[angle=0,width=14cm]{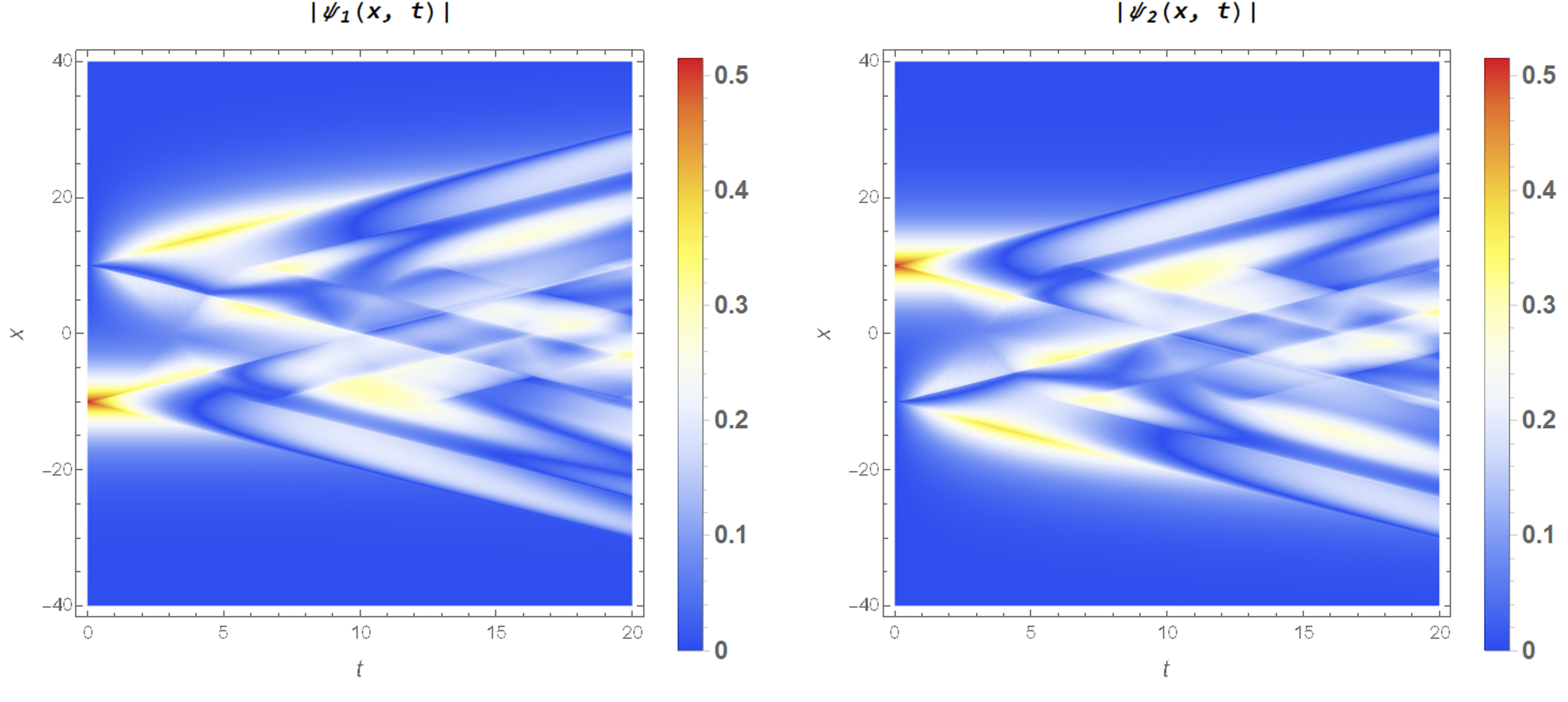}
\caption{The upper (left) and lower (right) components of the wave function,
for the coupling $g=0.25$.}
\label{fig-abs0.25}
\end{figure}

\begin{figure}
\centering
\includegraphics[angle=0,width=14cm]{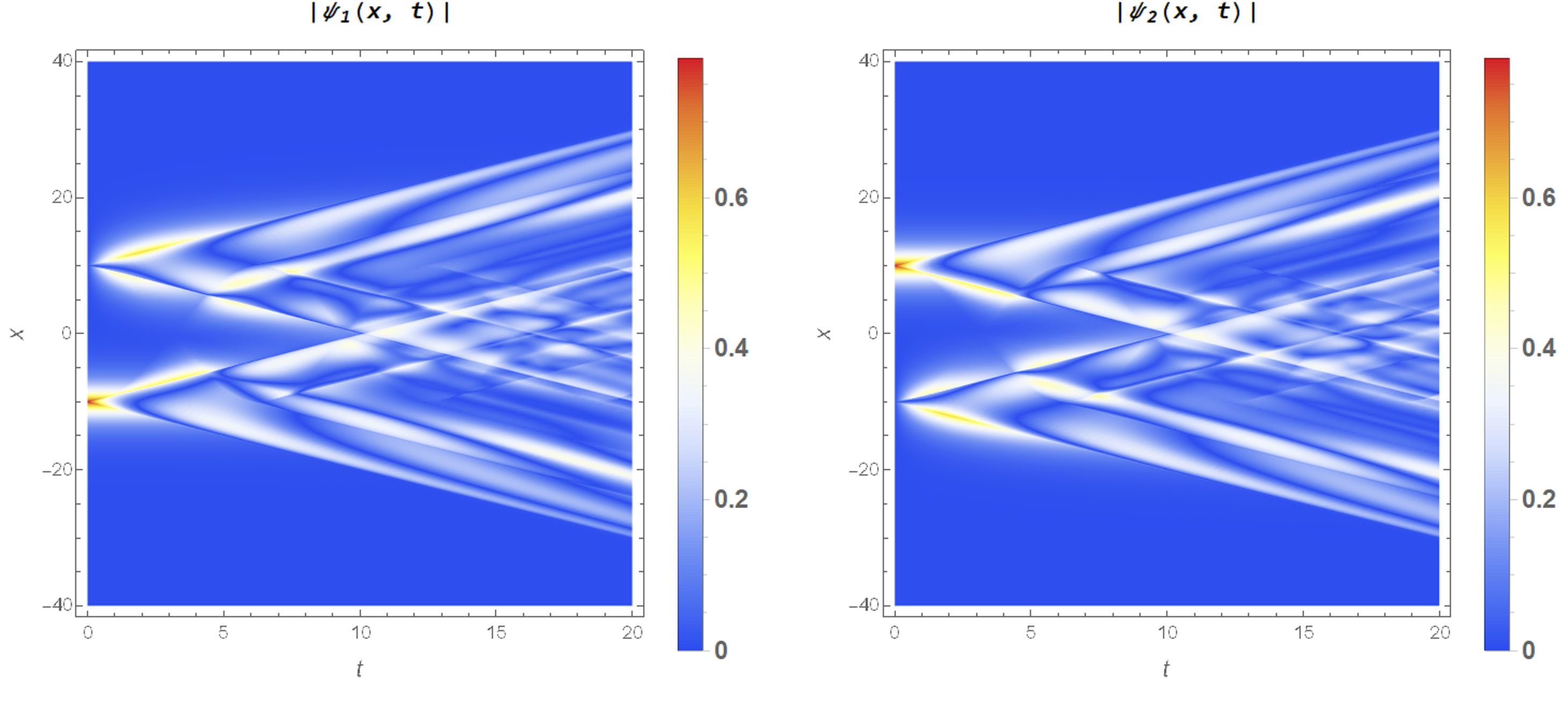}
\caption{The upper (left) and lower (right) components of the wave function,
for the coupling $g=0.50$.}
\label{fig-abs0.50}
\end{figure}

To deal with these complications, we integrated the evolution over an extended spatial domain, so that the
leading edge of the outgoing pulse, moving at the signal speed $c_{s}=1$, never reached the vicinity
of the boundary. The resulting wave functions were then simply cropped down to the region of
physical interest. Figures~\ref{fig-abs0.10} through~\ref{fig-abs0.50} show the results (plotted as
the absolute values of $\psi_{1}$ and $\psi_{2}$) for coupling values $g=0.10$, 0.25, and 0.50. The
initial states used in the calculations were negative-energy bound states of the
potential at time $t=0$---the almost-zero-energy modes of $V(x,0)$ that are symmetric combinations
of the localized modes around $v=10$ and $-v=-10$~\cite{ref-scott1}.

Studying these plots, a number of significant features may be observed. The parity symmetry is
clearly evident, since $\psi_{1}(x,t)$ and $\psi_{2}(x,t)$ are mirror images of each other across the line
$x=0$. (This provides an important ``sanity check'' of the computational results.) The diagonal
lines indicating the propagation of fronts are also very obvious. The slopes of the visible lines are,
as they should be, the signal velocities $\pm c_{s}$; the signal speed $c_{s}$ is
the phase velocity for the shortest-wavelength modes, which
is, according to~\eqref{eq-large-x}, simply the propagation speed of massless excitations in vacuum,
$c_{s}=1$.

\begin{figure}
\centering
\includegraphics[angle=0,width=14cm]{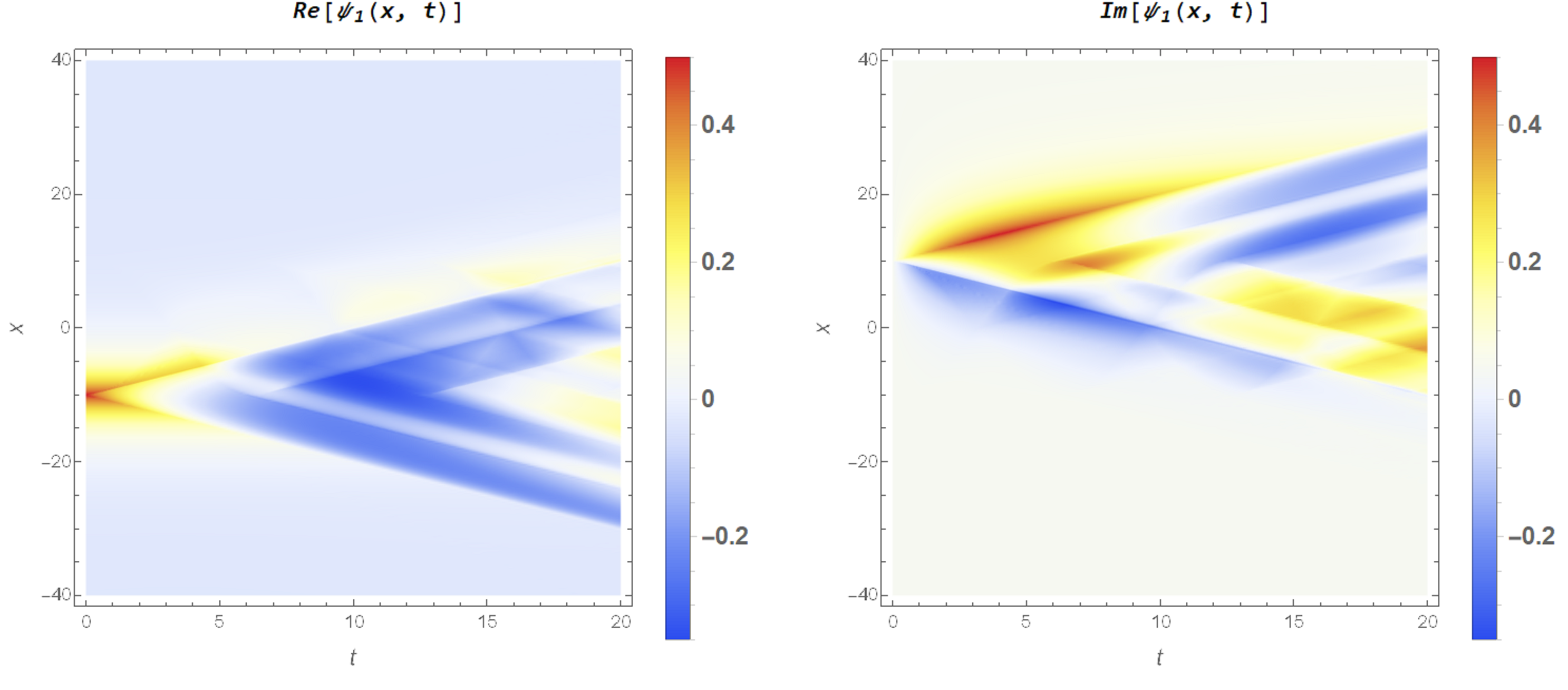}
\caption{Real and imaginary parts of $\psi_{1}$ for $g=0.25$.}
\label{fig-real-imaginary}
\end{figure}

Although initially $\psi_{1}$ is strongly localized around the kink and $\psi_{2}$ around the antikink,
the time-dependence of the background causes the solution around $x_{K}=-v=-10$ to acquire a $\psi_{2}$
component fairly quickly. The kink and antikink in~\eqref{eq-kinks} start from rest, but it does not
take long for their motion to couple the two wave function components together.
As a result, over most of the integration time interval, $\psi_{1}$ and $\psi_{2}$ are
comparable everywhere in the spatial domain. From each of the regions around $x=\pm v$ where the wave
functions are initially localized, each of $\psi_{1}$ and $\psi_{2}$ shows an pattern of expansion
at the speed $c_{s}$.
At short times, this may look like oscillation-stimulated tunneling from the bound state to continuum
states. However, the outflow from the soliton regions continues indefinitely; in particular, there continues
to be outgoing radiation after the oscillation amplitude of $|\psi|^{2}$ in the immediate
vicinity of the domain walls has fallen to be no
larger than the typical values elsewhere inside the light cone.
After a time $v/c_{s}=10$, the expanding cones overlap, and there
is an interference region centered on $x=0$. Interestingly, as a consequence of the way the
equations~(\ref{eq-psi1''}--\ref{eq-psi2''}) are coupled (with the coupling terms having explicit factors of $i$),
the contribution to $\psi_{2}$ in the initial vicinity of $\psi_{1}$ (and vice versa) is mostly
imaginary, while $\psi_{1}$ in that region remains predominantly real---at least until the
expanding light cones based at $\pm v$ begin to overlap. This fact is evident from
figure~\ref{fig-real-imaginary}, which shows the typical behavior of the real and imaginary parts of
$\psi$.

\begin{figure}
\centering
\includegraphics[angle=0,width=7cm]{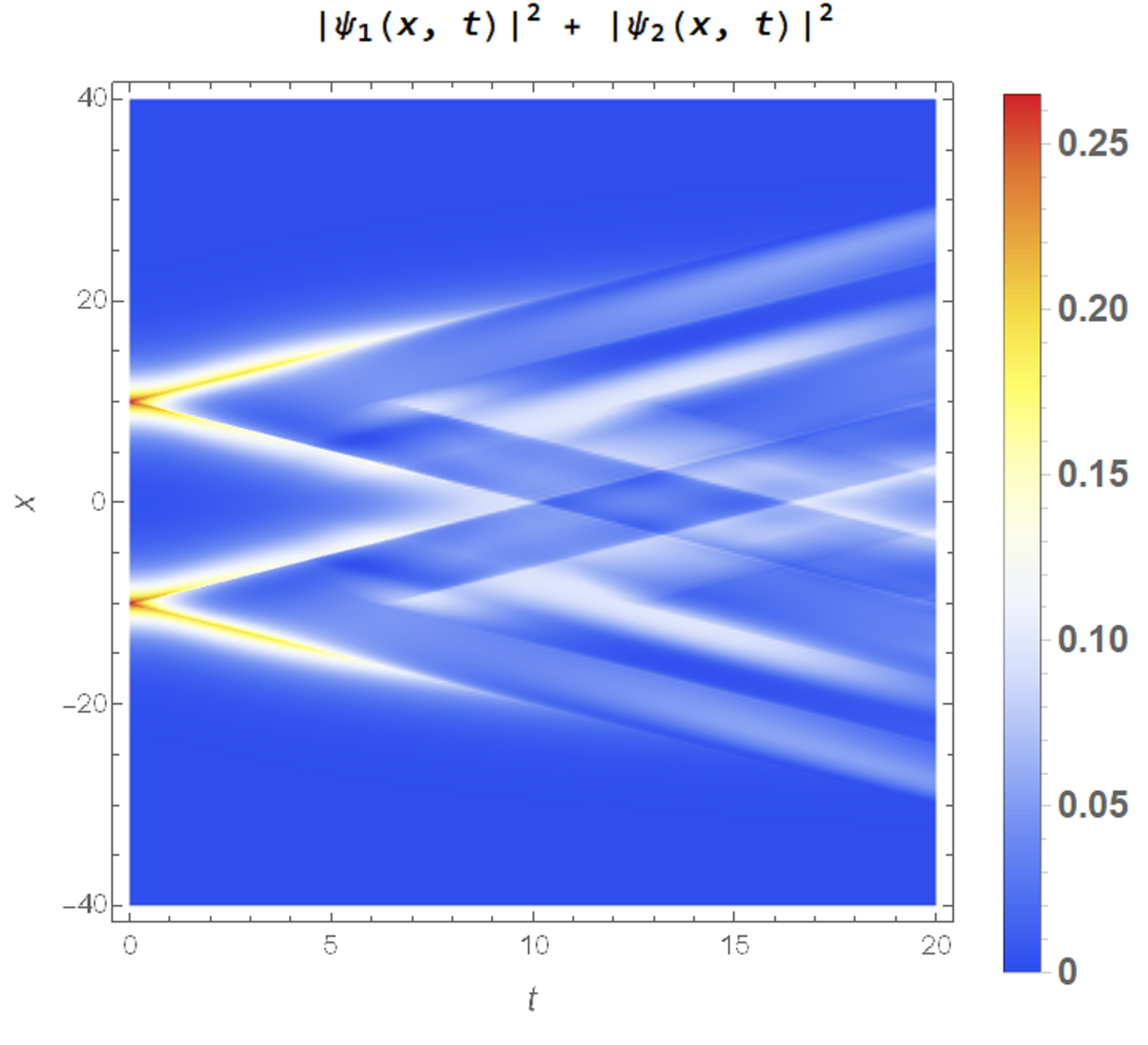}
\caption{Total particle density $\psi^{\dag}\psi$ for the solution with $g=0.25$.}
\label{fig-squared}
\end{figure}

In all the figures, but particularly in figure~\ref{fig-squared}, which shows the total
$\psi^{\dag}\psi=|\psi_{1}|^{2}+|\psi_{2}|^{2}$, the oscillation period $\pi/\omega=2.0\pi$ of the
domain wall potential~\eqref{eq-V-sgn} is also visible. There are quite evidently multiple pairs of
nested signal
cones. Although the greatest part of the flux lies on the primary shock fronts, enough
particle density remains near the initial positions at $v=\pm10$ for there to be partial
recurrences of the initial peaks after one oscillation period of the potential. 
These are indicated by the cusps in the plots that form the 
lying at $t$ coordinate $\pi/\omega$. There is another feature, of the same nature, at $t=2\pi/\omega$,
although it is markedly less pronounced, since relatively little particle density remains
anywhere near $\pm v$ after that much time has passed.

The partial recurrence---and thus the nested cones
that are visible to greater or lesser extents in all the
figures---is actually the most visually salient feature that is specific to a situation in which
the initial conditions represent a fermion wave function that is strongly peaked around one or both of
the domain walls. For initial states peaked elsewhere, the general light cone expansion looks not
especially dissimilar to what is shown in figures~\ref{fig-abs0.10}, \ref{fig-abs0.25},
and~\ref{fig-abs0.50}. Within each cone, there is an additional wave structure, visible as blue and
white striations in those three figures, with a wavelength that decreases with increasing $g$.

\section{Conclusions}

\label{sec-concl}

The paper has delved into the complex dynamics of fermion interactions in a time-dependent
breather-like background, specifically focusing on oscillating domain walls. Our analysis shows that,
despite the initial hopes for integrable behavior akin to the purely bosonic
sine-Gordon theory, the fermion modes exhibit unavoidable outward flux. This flux is driven
energetically by the oscillations of the scalar field, resulting in the eventual escape of
fermion-antifermion pairs to infinity, as illustrated in figures~\ref{fig-abs0.10}--\ref{fig-abs0.50}.

A key finding is that although fermion wave functions initially appear bound to the domain walls, over
time they do not remain localized, and indeed they do not remain significantly more localized than
arbitrary wave functions that are not closely approximated by bound states of the static forms
of the potentials. In either case, the oscillating backgrounds leads to the propagation of
fermionic states outward at speeds up to the speed of light, as demonstrated by the cone-shaped shock
fronts seen in the graphs (i.e. figure~\ref{fig-abs0.10} for $g=0.10$, figure~\ref{fig-abs0.25}
for $g=0.25$, and figure~\ref{fig-abs0.50} for $g=0.50$).
These graphs clearly depict the expansion of the fermionic excitations beyond the domain wall regions,
showing the spread of interacting
fermion wave functions across space. This is rather unlike the behavior often seen
in the purely bosonic sectors of higher-order non-integrable models~\cite{ref-khare,ref-bazeia2},
in which the effective potentials
between the solitary waves in $1+1$ dimensions~\cite{ref-christov1,ref-mello} tend to remain well localized, even
when the individual kinks and antikinks possess fat tails~\cite{ref-christov2}.

Numerical simulations show that even though some particle density remains localized near the initial positions
of the domain walls after one oscillation period (visible as partial recurrences in
figure~\ref{fig-squared}), the overall
trend is a wave of flux moving toward spatial infinity. This feature underscores the limitations of
approximating such systems as fully integrable, as the outward particle flux becomes a significant factor
in the system's long-term behavior. Moreover, after a sharp falloff in the fermion field amplitude at small
times, the oscillating fermion fields do not damp to zero at large times. Instead, their amplitudes remain
finite and comparable throughout the light cone interiors, as the motion of the kink-antikink pair continues
to pump energy into the fermion sector.

Further computational challenges arise due to the difficulty in imposing stable boundary conditions,
particularly in light of the fermionic modes propagating outward. Attempts to impose radiation-like boundary
conditions encountered
confounding issues, as damping methods applied to the Dirac equation led to standing waves
rebounding between the integration domain boundaries. This behavior is illustrated by the numerical
instabilities at the boundaries that arise over extended integration times, as discussed in
section~\ref{sec-computations}.
% XXX
%It is shown that for potentials in which the minima are Morse critical points, the interaction force decrease
%exponentially with the distance between the kink and the anti kink~\cite{ref-gonzalez}. However, in this case
%the defined potential does not possess true minima points due to its discontinuous, step-like form. The
%discontinuity at $x = \pm v\cos{\omega t}$ separates two regions of constant potential, rather than creating
%any local minima where particles might ``settle.''
%The kink-antikink force can also be calculated using uniform a ansatz for a coherently accelerating
%kink~\cite{ref-manton1}. However, this can be done for potentials that lead to a quartic minimum. In our case,
%the potetntial does not demand such a feature and the kink does not have a tail. So, more complex calculations
%are needed to calculate the force between the domain boundaries~\cite{ref-manton2}.

In addition to the outward flux of fermionic modes, our results also reveal interesting
quasiperiodic behavior
of the solutions, particularly around the time value of $\pi/\omega$, which corresponds to half the
oscillation period of the moving domain walls and one full
oscillation period of potential
$V(x,t)$. This periodicity is clearly reflected in the recurrence of fermion
density peaks at the initial positions of the domain walls, as seen in
figures~\ref{fig-abs0.10}--\ref{fig-squared}.
After each oscillation
cycle, the particle densities exhibit partial recurrences, although with diminishing intensity, as the fermion
modes continue to propagate outward.

In figure~\ref{fig-squared},
which shows the total particle density $\psi^{\dag}\psi$ for $g=0.25$, two distinct sets of
nested cones are clearly visible (and a third one less clearly). These cones reflect the periodic return of
fermion density near $x=\pm v=\pm 10$, which aligns with the oscillation period $\pi/\omega$. The first
recurrence is quite pronounced, with the peaks in particle density forming sharp cusps around $t=\pi/\omega$,
indicating that a significant portion of the fermion density remains bound near the domain walls after one
period. By $t=2\pi/\omega$, a second, less pronounced recurrence is visible, suggesting that most of the
particle density has dissipated away by this time, but a small amount still remains in the vicinity of the original
domain wall positions.

The oscillatory nature of the scalar potential, with period $\pi/\omega$, governs this periodic recurrence.
As the domain walls oscillate, they periodically couple and decouple the fermion wave functions, leading to
these recurrences of fermion densities after each full oscillation cycle. However, over time, the amplitude
of these recurrences decreases as more fermion density escapes toward spatial infinity. This feature, seen
in the graphs, underscores the non-adiabatic nature of the system, where the periodic motion of the domain
walls continuously injects energy into the fermionic modes, driving particle production and flux.

Overall, this analysis concludes that the sine-Gordon-like breather background, when coupled to fermions, does
not generally
support steady-state bound fermion solutions. Instead, the oscillations induce fermion particle production
and flux propagation, suggesting that the fermion-soliton system does not maintain the integrability and
stability expected from the purely bosonic sine-Gordon theory. It may be possible in the future to test
these theoretical results, in real systems with topological configurations of boson fields coupled to fermionic
particles.

\section*{Acknowledgments}

B. A. is grateful to the late R. Jackiw for many years of helpful discussions and, in particular, introducing
him to the topics of solitary waves and fermion fractionalization. The authors also appreciate the
assistance of S. Crittenden with the numerical solution of the Dirac equation.

\end{document}